\documentclass[aps,prl,superscriptaddress,twocolumn]{revtex4-2}

\pdfoutput=1 

\usepackage{graphicx}
\usepackage{siunitx}
\usepackage{amsmath}
\usepackage{xspace}
\usepackage[unicode=true,
  linktocpage,colorlinks=true,
  linkbordercolor={0.5 0.5 1},
  citebordercolor={0.5 1 0.5},
  linkcolor=blue]{hyperref}

\newcounter{superequation} 
\makeatletter
\@addtoreset{equation}{superequation}
\makeatother

\def\k{\mathbf{k}}
\def\q{\mathbf{q}}

\def\v{\mathbf{v}}
\def\fd{f^{(0)}}

\newcommand{\Bog}{\ensuremath{\rm B_{1g}}\xspace}
\newcommand{\Btg}{\ensuremath{\rm B_{2g}}\xspace}
\newcommand{\Tc}{\ensuremath{T_{\rm c}}\xspace}
\newcommand{\Tcs}{\ensuremath{T_{\rm c}}'s\xspace}

\def\bea{\begin{eqnarray}} \def\eea{\end{eqnarray}}

\newcommand{\SRO}{Sr$_2$RuO$_4$\xspace}

\usepackage[normalem]{ulem}

\usepackage{mathtools}

\DeclarePairedDelimiter\ket{\lvert}{\rangle}

\begin{document}

\def\titlename{Multipolar Fermi Surface Deformations in \SRO Probed by Resistivity and Sound Attenuation: A Window into Electron Viscosity and the Collision Operator}
\title{\titlename}

\author{Davis Thuillier}
\affiliation{Department of Physics and Astronomy, University of California, Irvine, Irvine, CA 92697, USA}
\author{Sayak Ghosh}
\affiliation{Geballe Laboratory for Advanced Materials, Stanford University, Stanford, CA, USA}
\affiliation{Department of Applied Physics, Stanford University, Stanford, CA, USA}
\author{B.~J.~Ramshaw}
\email{bradramshaw@cornell.edu}
\affiliation{Laboratory of Atomic and Solid State Physics, Cornell University, Ithaca, NY, USA}
\affiliation{Canadian Institute for Advanced Research, Toronto, Ontario, Canada}
\affiliation{These authors contributed equally to this work.}
\author{Thomas Scaffidi}
\email{tscaffid@uci.edu}
\affiliation{Department of Physics and Astronomy, University of California, Irvine, Irvine, CA 92697, USA}
\affiliation{These authors contributed equally to this work.}

\date{\today}

\begin{abstract}
Recent developments in electron hydrodynamics have demonstrated the importance of considering the full structure of the electron-electron scattering operator, which encodes a sequence of lifetimes, one for each component of the Fermi surface deformation in a multipolar expansion. In this context, the dipolar lifetime is measured by resistivity, whereas the quadrupolar component probes the viscosity and can be measured in the bulk via sound attenuation.
We introduce a framework to extract the collision operator of an arbitrary metal by combining resistivity and sound attenuation measurements with a realistic calculation of the scattering operator that includes multiband and Umklapp effects.
The collision operator allows for the prediction of a plethora of properties, including the non-local conductivity, and can be used to predict hydrodynamic behavior for bulk metals.
As a first application, we apply this framework to \SRO in a temperature range where electron-electron scattering is dominant. We find quantitative agreement between our model and the temperature dependence of both the resistivity and the sound attenuation, we find the quadrupolar (\Bog) relaxation rate to be 30\% higher than the dipolar one due to the presence of hot spots on the $\gamma$ band, and we predict a strongly anisotropic viscosity arising from the $\alpha$ and $\beta$ bands.
\end{abstract}

\maketitle

Particle flow is hydrodynamic when collisions between particles conserve energy, momentum, and particle number \cite{LandauLifshitz1987}. This is the case for liquid water, but for electrons in a metal, collisions with the lattice relax momentum and the flow is typically diffusive (i.e. Ohmic). Gurzhi~\cite{Gurzhi1968} showed that hydrodynamic effects should contribute to the electrical resistivity when momentum-conserving (i.e. non-Umklapp) electron-electron collisions dominate over other types of scattering. Under this condition, the electron viscosity---which controls the diffusion of momentum in the electron fluid---can become a relevant quantity in electric transport measurements \cite{Gurzhi1968,TomadinEtAl2014, TorreEtAl2015, LevitovFalkovich2016,   Zaanen2016, LucasHartnoll2018, LucasFong2018, Svintsov2018, ShytovEtAl2018, HuangLucas2021, NarozhnyEtAl2017, Narozhny2019, LevchenkoSchmalian2020, FritzScaffidi2024}. 

\begin{figure*}[t!]
\includegraphics[width=2\columnwidth]{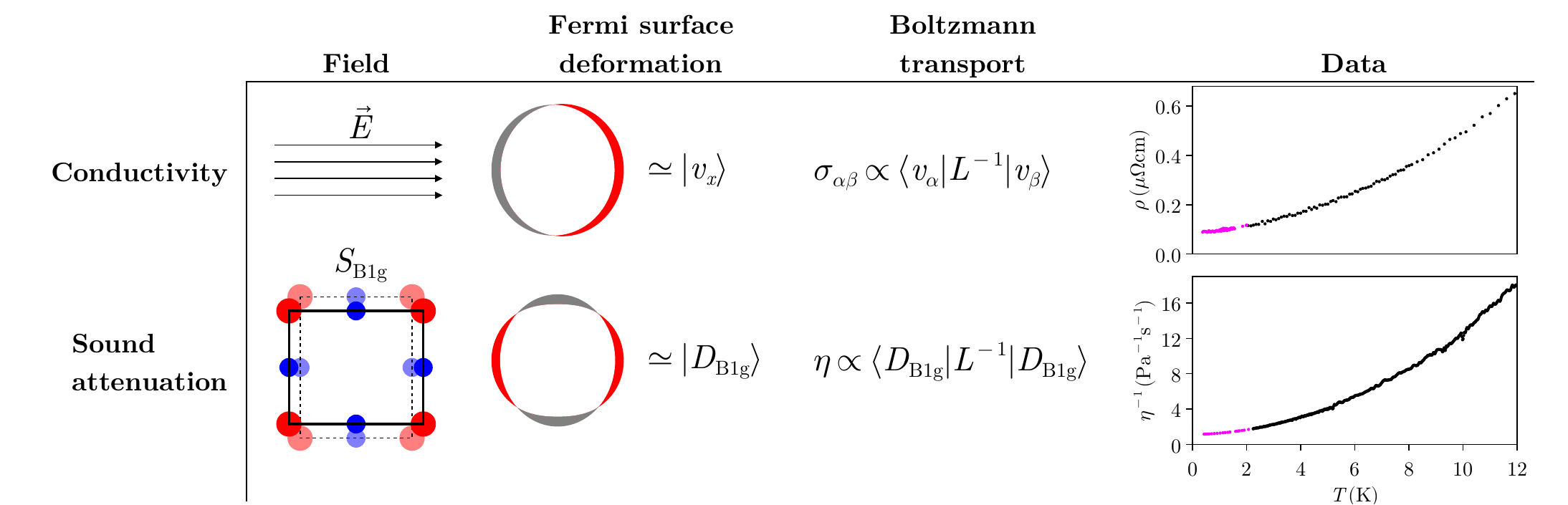}
\caption{\label{fig:fs_cartoons} \textbf{Comparison of electrical conductivity with sound attenuation.} The Fermi surface deforms under an electric field for conductivity, and a strain field for sound attenuation (red and gray regions correspond to population and depopulation, respectively). The Fermi surface deformation relevant to conductivity, $\sigma$, is proportional to the Fermi velocity $v_\alpha(\mathbf{k})$ and is thus dipolar; in contrast, the Fermi surface deformation relevant to sound attenuation, $\eta$, is determined by the deformation potential, $D_{\alpha\beta}(\mathbf{k})$, and is quadrupolar. Both conductivity and sound attenuation are calculated by evaluating the expectation value of the inverse collision operator, $L^{-1}$. The final column shows measurements of the resistivity (reproduced from \citet{Lupien2002}) and the inverse sound attenuation of \SRO: both quantities show Fermi liquid $T^2$ scaling. Data shown in pink are reproduced from \citet{LupienEtAl2001} and taken in a 1.5 T magnetic field to suppress the superconducting transition.}
\end{figure*}

Electron viscosity, however, has no impact on \textit{bulk} electric transport. Instead, it only enters in \emph{size-restricted} transport, for which the boundary of the device ultimately acts as the dominant momentum ``sink'' in the system \cite{YuEtAl1984, Black1980, MolenkampdeJong1994,BandurinEtAl2016, MollEtAl2016, Alekseev2016, KrishnaKumarEtAl2017, Scaffidi2017, HolderEtAl2019, ShavitEtAl2019,  SulpizioEtAl2019, BerdyuginEtAl2019,  SternEtAl2022, Aharon-SteinbergEtAl2022, KumarEtAl2022,KiselevSchmalian2019, Lucas2017, MoessnerEtAl2019,NazaryanLevitov2024, AlekseevSemina2018}. The reason why the viscosity does not contribute to bulk electric transport is fundamental: conductivity and viscosity probe the relaxation of different deformation modes of the Fermi surface (see \autoref{fig:fs_cartoons}).
Specifically, the bulk conductivity probes the relaxation of a dipolar Fermi surface (FS) deformation that is proportional to the Fermi velocity, e.g. $v_x \propto \cos(\theta)$ for a circular FS, with $\theta$ the angle around the FS.
By contrast, the electron viscosity is sensitive to the relaxation of quadrupolar FS deformations, 
e.g. varying as $\cos(2 \theta)$. 

Quadrupolar FS deformations are orthogonal to dipolar electrical currents in the bulk due their distinct symmetries. Viscosity can, however, contribute to electric resistance in size-restricted samples because translation symmetry breaking at the sample boundary generates quadrupolar (and higher-order multipolar) FS deformations ~\cite{Gurzhi1968, MolenkampdeJong1994}. These deformations clearly depend on details of how electrons interact with sample boundaries, and thus a more direct way to probe electron viscosity would be to generate \emph{bulk} quadrupolar FS deformations: these are the deformations generated by sound waves.

Schematically, sound waves traveling through a metal distort the lattice and change the electronic band structure through the deformation potentials \cite{Holstein1959,KhanAllen1987}, leading to quadrupolar deformations of the Fermi surface, with no net current (e.g. $D_{\Bog}\propto \cos(2 \theta)$, see \autoref{fig:fs_cartoons}).
At low temperatures, where phonon-phonon and phonon-dislocation scattering are weak (typically below about 20 K), and in the limit of long sound wavelength (compared to the electronic mean free path), sound attenuation
is dominated by the equilibration of the conduction electrons to the deformed lattice potential~\cite{Sorbello1976, KhanAllen1987}. In keeping with the conventions of the ultrasonic literature, this long-wavelength regime is referred to as the ``hydrodynamic'' regime \cite{KhanAllen1987}, and the attenuation coefficient $\alpha$ is typically expressed as an ``acoustic viscosity'': $\eta =\alpha \rho_m v_s/q^2$, where $\rho_m$ is the material density, $v_s$ is the sound velocity, and $q$ is the sound wavevector. 
(We will discuss below the difference between acoustic viscosity and the ``transport viscosity'' discussed in the context of electron hydrodynamics~\cite{FritzScaffidi2024}).

Consequently, the acoustic viscosity is proportional to the relaxation time for quadrupolar modes ($\eta \propto \tau_2$), whereas the conductivity is proportional to the transport mean free time ($\sigma \propto \tau_1$) that measures the relaxation of dipolar modes (See \autoref{fig:fs_cartoons}). Prior studies in elemental metals \cite{Steinberg,Bhatia,RiceSham1970, SathishEtAl1986, KolouchMcCarthy1965, Filson1959} have focused on the ratio $\tau_2/\tau_1$, but only in a regime for which either electron-impurity or electron-phonon scattering dominate~\cite{Kaveh01011984}.
Here, we are interested in the temperature regime that is dominated by electron-electron scattering, for which a parametric difference between the two rates is possible because the relaxation of currents requires Umklapp scattering~\cite{Ziman2001}, whereas the acoustic attenuation does not.

More generally, conductivity and viscosity probe the first two modes in a ``Fermi surface harmonics'' expansion (generalizing cylindrical or spherical harmonics to the case of an arbitrary FS shape) that encodes a hierarchy of timescales $\tau_l$ (with $l$ a generalized mode index) over which multipolar FS deformations relax.
Based on this fact, we propose to combine measurements of electrical conductivity and acoustic viscosity with theoretical calculations in order to construct the entire \emph{collision operator}, which is the central object in studies of novel regimes of transport~\cite{Callaway1959,MolenkampdeJong1994,LedwithEtAl2019,LedwithEtAl2019a, BakerEtAl2024a, HofmannDasSarma2022, HongEtAl2020, HofmannDasSarma2022, HofmannGran2022}.

The fine features of the collision operator were neglected for a long time due to the widespread use of the single relaxation time approximation (RTA), for which $\tau_l = \tau $ for all $l$.
This approximation was subsequently improved through the use of a two-rate model, for which the hydrodynamic regime is reached when $\tau_2^{-1}/\tau_1^{-1} \gg 1$~\cite{abrikosov,Callaway1959,MolenkampdeJong1994}, and more recently through the discovery of a tomographic regime for 2D metals~\cite{LedwithEtAl2019, LedwithEtAl2019a, HongEtAl2020, HofmannDasSarma2022, HofmannGran2022}.
However, even these more recent works have typically used simplistic models when calculating the collision operator (e.g. isotropic Fermi surfaces, no Umklapp)~\cite{abrikosov, SYKES19701,MolenkampdeJong1994, Scaffidi2017, HolderEtAl2019, HofmannDasSarma2022, HofmannGran2022, LedwithEtAl2019, Svintsov2018, ShytovEtAl2018, HuangLucas2021}.
While this might be justified in low-density conductors like two-dimensional electron gases~\cite{MolenkampdeJong1994, AhnDasSarma2022, GuptaEtAl2021} or graphene \cite{LevitovFalkovich2016, NarozhnyEtAl2017, Narozhny2019}, it is not applicable to many other candidate materials. A calculation of the scattering operator for realistic band structures is thus needed if we intend to find new materials with hydrodynamic regimes, and more generally if we are to understand the fundamental properties of scattering in metals, strange or not~\cite{HartnollMackenzie2022,Behnia1,Sun}.

As a proof of principle, we apply this framework to \SRO---a well-characterized Fermi liquid with moderately strong electron-electron interactions~\cite{RevModPhys.75.657,MackenzieEtAl2017,BergemannEtAl2003, HermanEtAl2019, HicksEtAl2014,Steppke,Barber2018,BarberEtAl2019,Sunko2019,Li2022,Chronister2022,NoadEtAl2023,Yang}.
We are interested in a temperature regime below $~\sim$ 12 K~\footnote{Above 12 K, phonon-phonon and phonon-dislocation scattering contribute significantly to the acoustic attenuation} for which both resistivity and inverse \Bog viscosity follow conventional Fermi liquid scaling (i.e. grow quadratically with temperature, see \autoref{fig:fs_cartoons} ~\footnote{ Note earlier work on \SRO focused on transport at higher temperatures within Dynamical Mean Field Theory~\cite{DengEtAl2016,AbramovitchEtAl2023}}). 

\begin{figure}[t!]
\includegraphics[width = 0.68\columnwidth]{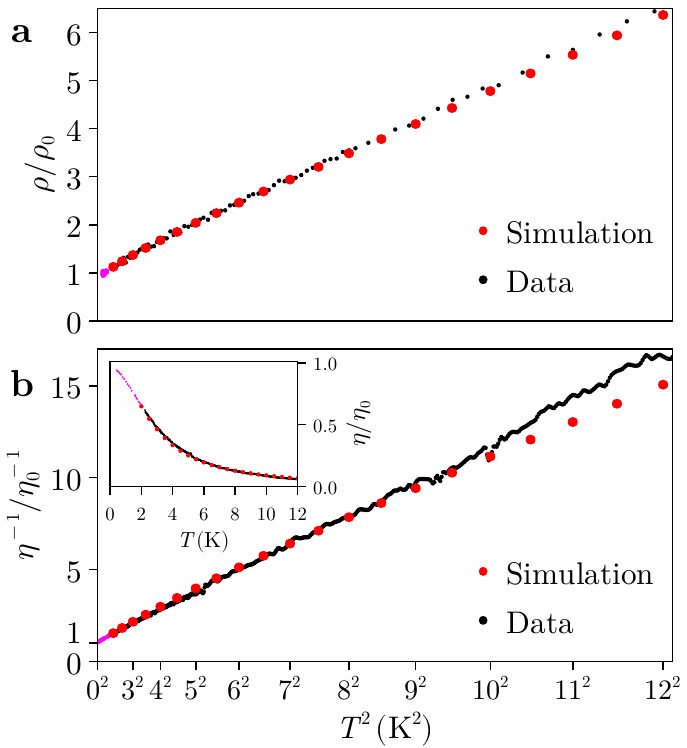}

\includegraphics[width = 0.7\columnwidth]{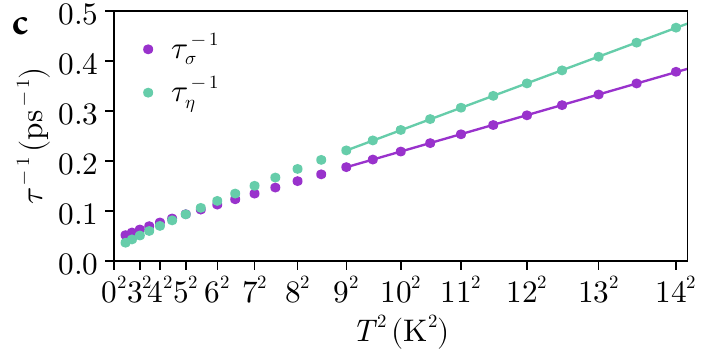}
\caption{\label{fig:res_vis_fits} \textbf{(a),(b) Comparison of experiment and theory for the resistivity and inverse viscosity as a function of temperature.} Two free parameters were adjusted to match $\rho(T)$; those same two parameters are then used to calculate the normalized inverse viscosity $\eta^{-1}/\eta_0^{-1}$, which shows excellent agreement with experiment. Above 10 K, the signal of the electronic contribution to sound attenuation becomes small compared to background, as shown in the inset, and is less reliable. Resistivity data from \cite{Lupien2002}. 
Data shown in pink are reproduced from \citet{LupienEtAl2001} and taken in a 1.5 T magnetic field to suppress the superconducting transition. \textbf{(c) Effective lifetimes extracted from the conductivity and viscosity.} Lines are fits to $A + B T^2$.}
\end{figure}

We find the following fits to the experimental data shown in \autoref{fig:res_vis_fits}:
\begin{equation}
    \begin{aligned}
\rho/\rho_0  = 1 + B_\rho T^2 \text{ and } 
\eta^{-1}/\eta^{-1}_0  = 1 + B_{\eta^{-1}} T^2
\end{aligned}
\end{equation}
with $B_\rho = 0.035 /K^2$ and $B_{\eta^{-1}} = 0.089 / K^2$.

The fact that $B_\rho$ and $B_{\eta^{-1}}$ differ by a factor of $\sim 2.5$ is a clear indication of the failure of the single RTA, and provides a very stringent consistency check on any theoretical model for electron scattering in this material.
The remainder of this Letter will demonstrate how a state-of-the-art numerical calculation of the collision operator for a realistic band structure illuminates the physics behind the difference of these prefactors.
We then use the extracted collision operator to predict non-local transport properties.

Before proceeding with the analysis, we first provide details of the acoustic attenuation measurements (see also End Matter). To access the long-wavelength limit of sound attenuation in \SRO, we used resonant ultrasound spectroscopy (RUS). Following ~\citet{GhoshEtAl2021,GhoshEtAl2022}, we measured all six elastic moduli and their respective attenuation coefficients, from 1.2 K to 12 K, of a single-crystal \SRO sample with a $T_c$ of 1.43 K. Here, we focus on the \Bog viscosity because it is particularly large in \SRO due to the proximity of the $\gamma$ Fermi surface to the van Hove points at the edge of the Brillouin zone. This produces an enhanced density of states along the (100) and (010) directions, for which the \Bog deformation potential is maximal~\cite{WalkerEtAl2001}.
By comparison, the \Btg viscosity does not even exceed the background in our experiment \cite{GhoshEtAl2022}.

We take resistivity data from \citet{Lupien2002}, measured on a sample with $\Tc \approx 1.42$ K. A sample cut from the same bulk crystal was used to perform pulse-echo sound attenuation measurements in \citet{LupienEtAl2001}\footnote{Note that the magnitude of the viscosity reported in \citet{LupienEtAl2001} is small by a factor of two---this was corrected in \citet{Lupien2002} (from Christian Lupien, private communication)}. We find that $\eta_{\Bog}$ measured by \citet{LupienEtAl2001} matches quantitatively with our RUS measurements. This ensures that the elastic scattering rate in our sample and the sample of \citet{LupienEtAl2001} is very similar---a fact also evidenced by the very similar \Tcs of our samples. We use the measurements from \citet{LupienEtAl2001} made in a magnetic field to extend the viscosity below \Tc (See End Matter for more details).

\emph{Collision operator}--- To calculate the conductivity and viscosity, we first construct the collision operator entering the Boltzmann equation: 
\bea
\partial_t f + \v  \cdot \nabla_r f + \mathbf{F} \cdot \nabla_k f = -L[f],
\eea
with $f$ the electron distribution function, $L=L_{ee} + L_\text{imp}$ the collision operator including both electron-electron (e-e) and electron-impurity (e-imp) scattering, $\v$ the Fermi velocity, $\k$ momentum, and $\mathbf{F}$ external forces.
We consider weak perturbations of the Fermi-Dirac distribution away from equilibrium of the form $f(\epsilon_\k) = \fd(\epsilon_\k) + \frac{-\partial\fd}{\partial\epsilon_\k} \chi(\k)$, 
where $\fd$ denotes the equilibrium Fermi-Dirac distribution, and $\chi(\k)$ describes the perturbation away from equilibrium.

The electron-electron collision operator  reads $  L_{ee}[\chi] \equiv  \int d\k' \mathcal{L}_{ee}(\k_1, \k^\prime) \chi(\k^\prime)$, with the kernel given by
\begin{equation}\label{eq:linearized_collision_operator}
    \begin{aligned}
         & \mathcal{L}_{ee}(\k_1, \k^\prime) = \left(\frac{\partial \fd}{\partial \epsilon_{\k_1}}\right)^{-1} \int_\text{BZ} \frac{d^2\k_2 d^2\k_3 d^2\k_4}{(2\pi)^6} \\
         &\times \Gamma(\k_1, \k_2, \k_3, \k_4) \fd(\k_1) \fd(\k_2)\overline\fd(\k_3) \overline\fd(\k_4) \\
         \times (\delta(&\k_1 - \k^\prime) + \delta(\k_2 - \k^\prime) - \delta(\k_3 - \k^\prime) - \delta(\k_4 - \k^\prime))
    \end{aligned}
\end{equation}
where $\Gamma$ denotes the rate for the scattering process $\k_1,\k_2 \rightarrow \k_3,\k_4$ and where $\overline\fd \equiv 1-\fd$.
We use a three-orbital, 2D Hubbard model for \SRO, in which case $\Gamma$ is given by a multi-orbital generalization of the formula $\Gamma = \frac{2\pi}{\hbar} U_{ee}^2 \delta(\k_1 + \k_2 - \k_3 - \k_4) \delta(\epsilon_{\k_1} + \epsilon_{\k_2} - \epsilon_{\k_3} - \epsilon_{\k_4}) $ (see Supplement Material (SM~\cite{supplement}) for details). 
\nocite{KhanEtAl1985,BurganovEtAl2016}
Extending the method of Refs.~\cite{Buhmann2013,HermanEtAl2019}, we numerically calculate the collision kernel of Eq.~\ref{eq:linearized_collision_operator} for a discretized annular region of $k$-space centered on each Fermi surface and of width proportional to $k_B T$, with 4884 patches in total (see SM~\cite{supplement} for more details). Generating the full collision operator at this resolution requires summing over $\sim 10^{11}$ scattering processes.

We also include the electron-impurity scattering operator $L_\text{imp}[\chi] \equiv  \int d\k_2 \mathcal{L}_\text{imp}(\k_1, \k_2) \chi(\k_2)$ with
\begin{equation}
    \mathcal{L}_\text{imp}(\k_1, \k_2) = \frac{2\pi}{\hbar} n_\text{imp} | \langle \k_1 | V_\text{imp} | \k_2 \rangle |^2 \delta(\epsilon_{\k_1} - \epsilon_{\k_2})
\end{equation}
where $n_\text{imp}$ is the impurity density and $V_\text{imp}$ measures the impurity potential strength.
We assume scattering on point-like impurities in the unitary limit~\footnote{In which case, $\langle \k_1 | V_\text{imp} | \k_2 \rangle$ is the $T_{\k_1,\k_2}$ matrix, strictly speaking\cite{Hewson_1993}.}, following previous work on \SRO~\cite{SuzukiEtAl2002, KikugawaEtAl2002}.
In this limit, and for an isotropic 2D band, $| \langle \k_1 | V_\text{imp} | \k_2 \rangle |^2 = 4 \hbar^2 v_F^2 / k_F^2$~\cite{Lapidus,SunkoEtAl2020}. 
We use a straightforward, multiband generalization of this formula, given in the SM~\cite{supplement}.

To calculate the full collision operator, we combine the electron-electron and electron-impurity scattering contributions and treat the energy scales for each contribution ($U_\text{ee}$ and $U_\text{imp}$) as our only two fit parameters:
\begin{equation}
    L = U_\text{ee}^2 \tilde{L}_\text{ee} + U^2_\text{imp} \tilde{L}_\text{imp},
\end{equation}
with $\tilde{L}_\text{ee} \equiv L_{ee} / U_{ee}^2$ and $\tilde{L}_\text{imp} \equiv L_\text{imp} / U_\text{imp}^2$, and where $U_\text{imp}^2 \equiv n_\text{imp} a^{-2} |\langle V_\text{imp} \rangle |_\text{av}^2 $ with $a$ being the lattice spacing and $|\langle V_\text{imp} \rangle |_\text{av}$ a band-averaged matrix element for electron-impurity scattering (see SM~\cite{supplement} for details).

With the collision operator in hand, the conductivity and viscosity are calculated as expectation values~\cite{Ziman2001,KhanAllen1987}:
\begin{equation}\label{eq:conductivityandviscosity}
    \begin{aligned}
    \sigma_{\alpha\beta} &= 2e^2 \langle v_\alpha | L^{-1} |v_\beta\rangle,\\
            \eta_{\alpha\beta\gamma\delta} &= 2 \langle D_{\alpha\beta} | L^{-1} |D_{\gamma\delta}\rangle
    \end{aligned}
\end{equation}
where $v_\alpha = \hbar^{-1}\nabla_{k_\alpha} \epsilon_\k$ is the Fermi velocity, $D_{\alpha\beta} = \frac{\partial \epsilon_\k}{\partial S_{\alpha\beta}}$ is the deformation potential~\cite{KhanAllen1987}\footnote{The deformation potentials are calculated based on the strain dependence as inferred from uniaxial strain experiments~\cite{BarberEtAl2019, NoadEtAl2023}(see SM~\cite{supplement} for more details).}, and $\langle \psi | \chi \rangle \equiv \int_\text{BZ} \frac{d^2 \k}{(2\pi)^2} \frac{-\partial \fd}{\partial \epsilon_\k} \psi^*(\k) \chi(\k)$. The factors of 2 are due to spin.
As mentioned above, we will focus on the \Bog viscosity which we call $\eta$ for short, and for which $| D_{\Bog} \rangle = \frac12 |D_{xx}\rangle - \frac12 |D_{yy}\rangle$.

We first fit our two parameters to match the measured resistivity $\rho(T)$ (\autoref{fig:res_vis_fits}a) and obtain a good fit for $U_\text{ee} = \SI{0.074}{\electronvolt}$ and $U_\text{imp} = \SI{4.8e-4}{\electronvolt}$.
Using the exact same parameters, we then predict the \Bog viscosity and find good agreement with experiments (\autoref{fig:res_vis_fits}b). 
That the calculation predicts good agreement with experimental viscosity---using the parameters inferred from the resistivity---is nontrivial, since the experimental data clearly deviates from the single RTA as discussed above: over the temperature range we consider, the resistivity increases by a factor of $6$ whereas the viscosity increases by a factor of $16$. If the single RTA was valid, then both quantities would have the same relative increase.

\emph{Effective relaxation times}---To highlight the breakdown of the single RTA, it is useful to define effective relaxation times corresponding to each quantity, analogous to $\tau_1$ and $\tau_2$ defined in the introduction.
However, for an anisotropic, multi-band system like \SRO, $|v_\alpha\rangle$ and $|D_{\alpha\beta} \rangle$ are, in general, not eigenmodes of $L$, thus conductivity and viscosity cannot be associated with a single eigenvalue of $L$. Nevertheless, effective lifetimes corresponding to the conductivity and the (\Bog) viscosity can be defined based on \eqref{eq:conductivityandviscosity} such that ``generalized Drude formulas'' hold, namely $\sigma = 2 e^2 \langle v_x | v_x \rangle \tau_\sigma$ and $\eta^\text{ac} = 2 \langle  D_{\Bog} |  D_{\Bog}\rangle \tau_\eta$.

The breakdown of the single RTA is confirmed by comparing the effective lifetimes for each quantity, as shown in \autoref{fig:res_vis_fits}c. 
We see that both inverse lifetimes grow like $A+B T^2$ but with a prefactor $B$ that is 30\% larger for $\eta$ than for $\sigma$. 
This 30\% difference is attributed to ``hot spots''~\cite{Mousatov} on the $\gamma$ band. These are regions near the van Hove points along (100) and (010) where an increased density of states enhances scattering. Crucially, these hot spots are much more efficient at relaxing \Bog deformations than currents, explaining the observed difference in effective lifetimes (see SM~\cite{supplement} for more details).

\begin{figure}[t!]
\includegraphics[width = 0.8\columnwidth]{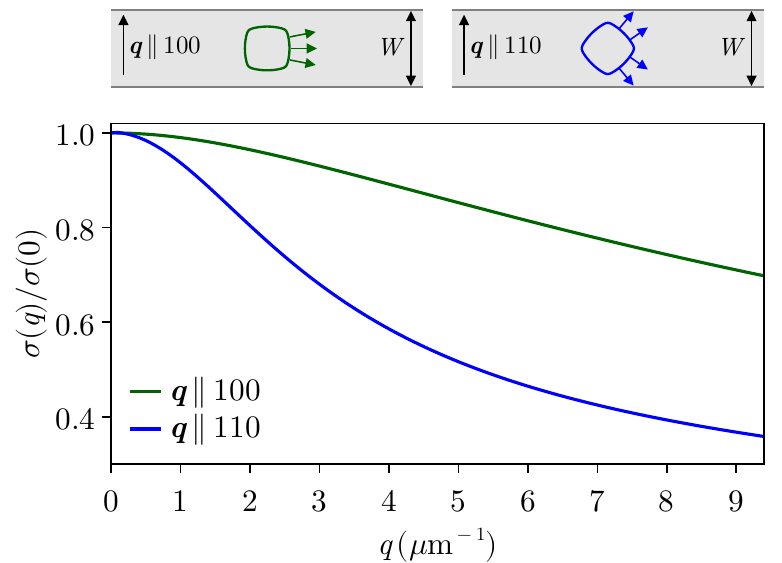}
\caption{\label{fig:nonlocal-conductivity}  \textbf{Non-local conductivity} calculated at $T = 14$ K. 
Top: schematic showing the $\beta$ FS of \SRO and selected Fermi velocities within an effective ``channel'', for $\q$ along 100 and 110.
}
\end{figure}

\emph{Non-local conductivity}---
There is considerable interest in predicting size-restricted conductivity because this experimental configuration has been used extensively as a probe of electron viscosity~\cite{FritzScaffidi2024}.
Now that we have an accurate collision operator, we could use it in a spatially-dependent Boltzmann calculation with appropriate boundary conditions for a given sample geometry. Such a calculation, however, would take us beyond the scope of this Letter.
Instead, we can calculate the closely related non-local conductivity $\sigma(\q)$~\cite{BakerEtAl2023, NazaryanLevitov2024}:
\begin{equation}\label{eq:nonlocal-conductivity}
    \sigma_{\alpha\beta}(\q) = 2 e^2 \langle v_\alpha | (L + i \q \cdot \mathbf{v})^{-1} | v_\beta \rangle.
\end{equation}
Because it probes the flow of electrons in response to an electric field with transverse spatial modulation (e.g. $E_x = E e^{i q y}$), the transverse non-local conductivity---which we will denote simply $\sigma(\q)$--- can be regarded as a proxy for the conductivity of electrons moving along $x$ in a channel whose width along $y$ is approximately half the electric field wavelength, i.e. $W \sim \pi/q$ (see~\autoref{fig:nonlocal-conductivity} top panel). 
An additional class of experiments sensitive to $\sigma(\q)$ is electromagnetic measurements that involve a spatially-varying electric field in bulk samples \cite{AgarwalEtAl2017a, KolkowitzEtAl2015}, e.g. using the skin effect \cite{ReuterEtAl1948, ForcellaEtAl2014, MatusEtAl2022, ValentinisEtAl2023}.

Our prediction for $\sigma(\q)$ at $T=14$~K along (100) and (110), shown in \autoref{fig:nonlocal-conductivity}, reveals a strong anisotropy, whereby $\sigma(\q)$ decays much more weakly for $\q$ along (100). In analogy with earlier work on systems with polygonal Fermi surfaces like PdCoO$_2$~\cite{Mackenzie2017,nandi_unconventional_2018,CookLucas,BachmannEtAl2022,BakerEtAl2023,BakerEtAl2024}, we attribute this anisotropy to the square-like Fermi surfaces of the $\alpha$ and $\beta$ bands, which lead to an ``easy direction'' for current propagation in narrow channels. This strong anisotropy provides a sharp experimental test of which bands govern non-local transport, revealing that the square-like $\alpha$ and $\beta$ bands dominate over the opposing influence of the nearly-circular $\gamma$ band. Furthermore, the decay of $\sigma(\q)$ with $q=|\q|$ allows for the definition of an anisotropic ``transport viscosity'' $\nu^\text{tr}$, following $\sigma(\q)/\sigma(0) = 1 - \nu^\text{tr}(\phi) \tau_\sigma q^2 + \mathcal{O}(q^4)$, where $\phi$ is the angle of $\q$ with the (100) crystallographic axis. This formula can be used to define a transport viscosity in any material, and is consistent with the Stokes-Ohm equation of electron hydrodynamics~\cite{Gurzhi1968,TorreEtAl2015,LevitovFalkovich2016} when the latter applies (see End Matter for more details on the difference between acoustic and transport viscosities, along with data on the temperature dependence of $\nu^\text{tr}$). The anisotropy of $\sigma(\q)$ is then reflected into a strongly anisotropic transport viscosity tensor with $\nu^\text{tr}_{\Bog} \gg \nu^\text{tr}_{\Btg}$, which was already predicted for electron-electron scattering at polygonal (in our case, square) Fermi surfaces and results from a long-lived imbalance mode~\cite{CookLucas}, the presence of which we confirmed numerically (see SM~\cite{supplement}). This result calls for more direct experimental probes of this mode and its resulting generalized hydrodynamics regime~\cite{CookLucas}.

In conclusion, we have presented a framework for constructing the full Boltzmann collision operator, paving the way for future investigations of diverse materials and properties---such as thermal~\cite{Jiang} and magnetotransport~\cite{Grissonanche}---that are highly sensitive to its detailed structure.

\begin{acknowledgments}
The work of D.T. and T.S. on the theoretical formalism and the numerical calculations was supported by the U.S. Department of Energy, Office of Science, Office of Basic Energy Sciences under Early Career Research Program Award Number DE-SC0025568. B.J.R and S.G. acknowledge support for building the experiment, collecting and analyzing the data, and writing the manuscript from the Office of Basic Energy Sciences of the United States Department of Energy under Early Career Research Program Award Number DE-SC0020143. We gratefully acknowledge Christian Lupien for providing the resistivity data below 2 K measured in a magnetic field, as well as providing information about their previously-published resistivity and viscosity data.
We gratefully acknowledge discussions with Avi Shragai, Aaron Hui, Andrew Mackenzie, Graham Baker, and Veronika Sunko.
Simulation codes are available on Github in the form of the package Ludwig.jl v0.1.0 \cite{Ludwig}.
\end{acknowledgments}

\bibliography{main.bib}

\onecolumngrid
\begin{center}
\ \vskip 0.2cm
{\large\bf End Matter}
\end{center}
\twocolumngrid

\renewcommand{\theequation}{A\arabic{equation}}
\stepcounter{superequation}

\emph{Experimental Methods}---
To access the long-wavelength limit of sound attenuation in \SRO, we used resonant ultrasound spectroscopy (RUS). We measured all six elastic moduli and their respective attenuation coefficients, from 1.2 K to 12 K, of a single-crystal \SRO sample with a $T_c$ of 1.43 K. The measurement frequency is approximately 2 MHz. The details of the experimental procedure, including sample characterization, are given in \citet{GhoshEtAl2021}. The details of extracting the sound attenuation and viscosity from the resonance linewidths are given in \citet{GhoshEtAl2022}. 

We extend the sound attenuation data to temperatures below \Tc by including data with a 1.5 T in-plane magnetic field from \citet{LupienEtAl2001}. Note that, because the magnetic field is in the ruthenium-oxide plane, there is no orbital contribution to the sound attenuation (the sound-attenuation equivalent of magnetoresistance). This is demonstrated explicitly in Figure 5.19 of \citet{Lupien2002}, which shows that the \Bog sound attenuation is independent of in-plane magnetic field above $H_{\rm c2}$.

\begin{figure}
    \centering
    \includegraphics[width=0.7\linewidth]{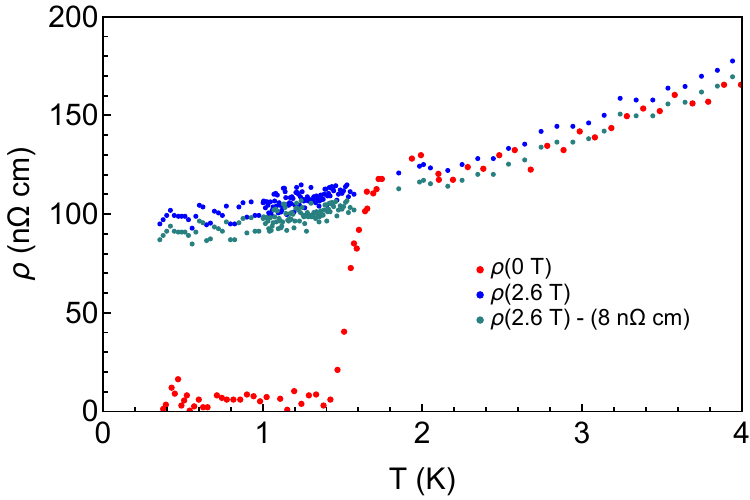}
    \caption{Resistivity data taken in zero field (red points) and in a 2.6 T field applied along the $c$ axis (blue points). The main text uses the zero-field resistivity data above 2.5 K, and the 2.6 T data below 2.5 K with a 8 n$\Omega\cdot$cm offset subtracted. }
    \label{fig:resist_subtract}
\end{figure}

We extend the resistivity data to temperatures below \Tc by including data taken in a 2.6 T out-of-plane magnetic field. Note that the resistivity obtained in this magnetic field configuration contains an orbital magnetoresistance contribution. \autoref{fig:resist_subtract} shows the raw resistance taken in zero magnetic field (red points), the raw resistance taken in a 2.6 T magnetic field (blue points), and the resistance taken in a 2.6 magnetic field with a 0.008 n$\Omega\cdot$cm offset subtracted (green points) so that it aligns with the zero-field resistivity above \Tc. 

\emph{Difference between acoustic and transport viscosity---}\label{end-matter}
Here we highlight how two different definitions of viscosities appear in the context of acoustic attenuation and non-local transport. Although they both ultimately are written in terms of the collision operator and involve calculating expectation values with distribution functions in the same symmetry sector, the distributions involved are different in the two cases, leading in general to a parametric difference between them.
Schematically, we will show that the acoustic viscosity is given by
\bea
\label{Eq:acousticeta}
 \eta^\text{ac}_{\alpha\beta\gamma\delta} \sim \langle D_{\alpha\beta} | L^{-1} | D_{\gamma \delta}\rangle
 \eea
 with $D$ the deformation potential, whereas the transport viscosity reads
 \bea
 \label{Eq:transporteta}
 \eta^\text{tr}_{\alpha\beta\gamma\delta} \sim \langle m v_\alpha v_\beta | L^{-1} | m v_\gamma v_\delta \rangle
 \eea
 with $m$ an average effective mass defined below.
The acoustic and transport viscosities are thus related to expectation values of the collision operator for states given by $|D_{\alpha\beta}\rangle$ and $|m v_\alpha v_\beta\rangle$ respectively, which have the same units and the same symmetry, but can in general be parametrically different since $D$ is given by the derivative of the energy with strain, whereas $v$ is related to the derivative of the energy with $\k$.

A simple illustrative example is to consider a nearest-neighbor 1D hopping model for which the hopping amplitude depends on the strain $\epsilon$ as $t(\epsilon) = t (1 - \alpha \epsilon)$, with $\alpha$ a dimensionless number. In this case, one finds $D \propto t \alpha \cos(k_F)$ but $m v^2 \propto t k_F |\sin(k_F)|$. Clearly these two distributions have a different parametric dependence on the model parameters: e.g. when $k_F \simeq \pi/2$, $D$ is parametrically small whereas $m v^2$ is not.

In the case of \SRO, this could lead to qualitatively different behaviors for $\eta^\text{tr}_{\Bog}$ and $\eta^\text{ac}_{\Bog}$.
Both viscosities involve the relaxation time of a quadrupolar mode in the \Bog sector, but the shape of the mode is different: in the acoustic case it is the deformation potential $\ket{D_{\Bog}} \equiv \ket{\frac{\partial \epsilon_k}{  \partial  S_{\Bog} }}$ whereas in the transport case it is given in terms of Fermi velocities $\ket{\frac12 m(v_x^2 - v_y^2) }$.
Notably, at the van Hove points $v_x^2 - v_y^2$ vanishes whereas $D_{\Bog}$ does not, which would lead to drastically different behavior for $\eta^\text{tr}_{\Bog}$ and $\eta^\text{ac}_{\Bog}$ across the Lifshitz transition. This could be checked experimentally under uniaxial strain following earlier work~\cite{HicksEtAl2014,Steppke,Barber2018,BarberEtAl2019,HermanEtAl2019,Sunko2019,Mousatov,Li2022,Chronister2022,NoadEtAl2023,Stangier,Yang}.

Further, the deformation potentials depend on the strength of the electron-phonon coupling, which appears in a tight-binding model as dimensionless numbers giving the hopping dependence on strain, e.g. $\alpha$ in the 1D example above.
By contrast, $|m v_\alpha v_\beta\rangle$ is impervious to the electron-phonon coupling. For the simple example above, this means $\eta^\text{ac} \propto \alpha^2$, whereas $\eta^{tr}$ is independent of $\alpha$. Since $\alpha$ can in general be large (we used $\alpha \simeq 7$ in our \SRO model for nearest-neighbor hopping), this can lead to orders of magnitude difference between the two viscosities.

Let us now derive the formulas \ref{Eq:acousticeta} and \ref{Eq:transporteta} for the two viscosities.
First, in the main text we have given the following formula for the acoustic viscosity~\cite{KhanAllen1987}:
\begin{equation}\label{}
 \eta^\text{ac} =  \langle D | L^{-1} | D \rangle
\end{equation}
with $D_{\alpha \beta} = \partial E / \partial S_{\alpha \beta} $ the deformation potentials. (We drop factors of 2 for spin in this End Matter.)

\begin{figure}[t!]
    \centering
    \includegraphics[width=0.7\columnwidth]{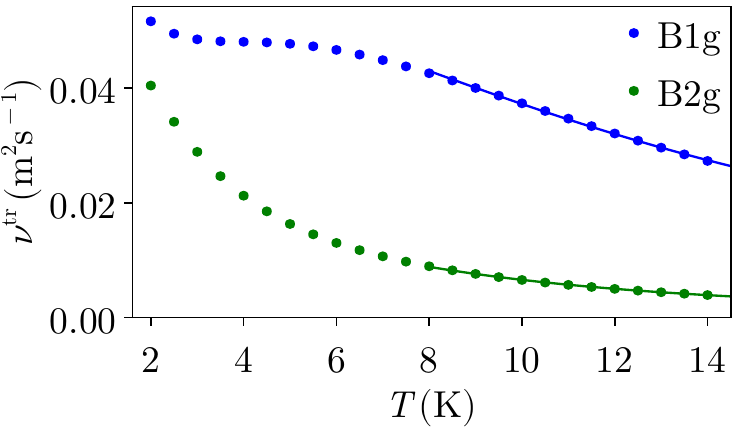}
    \caption{Transport viscosities calculated as a function of $T$. Lines are fitted to $(A+B T^2)^{-1}$.}
    \label{fig:transport-viscosities}
\end{figure}

As explained in the main text, a ``transport viscosity'' $\nu^\text{tr}$ can be defined based on non-local electric transport, through the small-$q$ decay of $\sigma(q)$:
\bea
\sigma(q)/\sigma(0) = 1 - \nu^\text{tr}(\phi) \tau_\sigma q^2 + \mathcal{O}(q^4)
\eea
where $\phi$ is the angle of $\q$ with the (100) crystallographic axis and $\tau_\sigma$ is the effective mean free time giving the bulk conductivity. 
By expanding
\begin{equation}
\sigma_{\alpha\beta}(\q) =  e^2 \langle v_\alpha | (L + i \q \cdot \mathbf{v})^{-1} | v_\beta \rangle
\end{equation}
in powers of $i \q \cdot \v$, one finds
\begin{equation}\label{Eq:threepowersofL}
\nu^\text{tr} = \frac{\langle v_\perp|L^{-1} v_\parallel L^{-1} v_\parallel L^{-1} | v_\perp\rangle }{\langle v_\perp|L^{-1}|v_\perp\rangle^2 \langle v_\perp|v_\perp\rangle^{-1}}
\end{equation}
where $v_{\perp}$ and $v_{\parallel}$ are respectively perpendicular and parallel to $\q$.
Using this formula, we calculated two components of the transport viscosity: $\nu^\text{tr}_{\Bog} \equiv  \nu\left(\phi=\frac{\pi}{4}\right) $ and  $\nu^\text{tr}_{\Btg} \equiv  \nu\left(\phi=0\right)$.
As shown in Fig.~\ref{fig:transport-viscosities}, each component follows an $(A+BT^2)^{-1}$ law, as expected. One can also see that the anisotropy $\nu^\text{tr}_{\Bog} \gg \nu^\text{tr}_{\Btg}$ discussed in the main text is strongest at higher temperatures, at which electron-electron scattering dominates.

Finally, we note that the transport viscosity naturally appears as a kinematic viscosity $\nu^\text{tr}$ in units of meters squared per second. We will now define its dynamic version $\eta^\text{tr}$ in units of pascal-seconds, which makes the comparison with the acoustic viscosity more transparent.
The numerator in Eq.~\ref{Eq:threepowersofL} involves three factors of $L^{-1}$ and thus in general cannot be expressed in terms of an expectation value of a single factor of $L^{-1}$ as $\eta^\text{ac}$ could. 
It is however instructive to consider the case when $|v_\perp \rangle$ is close to an eigenvector of $L$ (which is exactly true for a circular FS), i.e. $L^{-1} | v_\perp \rangle \simeq \tau_\sigma |v_\perp \rangle$, in which case the formula above simplifies to
\begin{equation}\begin{aligned}\label{}
\nu^\text{tr} &\simeq \frac{\langle v_\perp v_\parallel |  L^{-1}  |v_\perp v_\parallel \rangle}{\langle v_\perp|v_\perp\rangle} 
\end{aligned}\end{equation}
Explicitly for the two principal directions, we then find:
\begin{equation}\begin{aligned}\label{eq:nu}
 \nu^\text{tr}_{\Bog} &\equiv  \nu\left(\phi=\frac{\pi}{4}\right)  \simeq \frac{\langle \frac12 (v_x^2 - v_y^2) |  L^{-1}  |\frac12 (v_x^2 - v_y^2) \rangle}{\langle v|v\rangle} \\
\nu^\text{tr}_{\Btg}   &\equiv  \nu(\phi=0)  \simeq \frac{\langle v_x v_y |  L^{-1}  |v_x v_y \rangle}{\langle v|v\rangle} 
\end{aligned}\end{equation}
with $\langle v|v\rangle \equiv \langle v_x|v_x\rangle = \langle v_y|v_y\rangle$.

Now, to obtain a dynamic viscosity, we add a factor of mass density $n m$ with $n$ the carrier density and $m \equiv n \langle v_\perp | v_{\perp}\rangle^{-1}$ the average effective mass, leading to
\bea
\eta^\text{tr} \simeq \langle m v_\perp v_\parallel |  L^{-1}  |m v_\perp v_\parallel \rangle
\eea
which takes the form given in \eqref{Eq:transporteta}. (One can check that the definition for the mass above gives $m = k_F / v_F$ for a circular Fermi surface).

\end{document}


\setlength{\belowcaptionskip}{0pt}

\begin{center}
    \large\textbf{Supplemental material to ``\textit{\titlename}''}
\end{center}

\appendix
\onecolumngrid
\setcounter{secnumdepth}{3}
\setcounter{figure}{0}
\setcounter{equation}{0}

\section{Definition of the high and low-frequency limits of sound attenuation}\label{se:limits}
Sound attenuation in the short-wavelength (``quantum'') limit is dominated by the direct  production of particle-hole pairs by ultrasonic phonons \cite{KhanAllen1987}. This limit is reached when the product of the sound wavevector $q$ and electron mean free path $l$ is of order 1. In \SRO, which has an electron mean free path of order 1 $\mu$m and a $B_{1g}$ shear sound velocity of 3 km/s, this crossover occurs at around 500 MHz. The resonant ultrasound data used in this manuscript were taken at $\approx 2$ MHz are therefore well below this limit and in the ``hydrodynamic'' regime of sound attenuation \cite{KhanAllen1987}.

Previous measurements of the $B_{1g}$ sound attenuation in Sr$_2$RuO$_4$, such as \citet{LupienEtAl2001}, were performed at higher frequencies using the pulse echo ultrasound technique. Therefore, they may contain sizable contributions from particle-hole pair production. We find, however, that at least for the \Bog viscosity, our resonant ultrasound measurements performed at $\approx 2$ MHz are in good quantitative agreement with those of \citet{LupienEtAl2001} measured by pulse-echo ultrasound at $\approx 40$ MHz.

\section{Scattering Rate\label{app:effective_rate}}

Our calculation of the collision operator is a generalization of the method employed in \cite{Buhmann2013} to a material with multiple bands. The electron-electron collision operator has been calculated previously for the $\gamma$ band of \SRO  in \citet{HermanEtAl2019} using this approach in order to consider how transport coefficients deviate from expected Fermi liquid behavior when a band undergoes a Lifschitz transition. However, for comparing the conductivity and the viscosity, it is necessary to capture how the $\alpha$ and $\beta$ contribute with a greater weight to the conductivity than the viscosity due to $D_{\Bog}$ being smaller on these bands compared to the $\gamma$ band. 

The low-energy physics of electrons in the $d$-orbitals of the ruthenium atoms in \SRO is modeled by the following Hamiltonian, 
\begin{align}\label{eq:hamiltonian}
    H &= H_0 + H_\text{int}\\
    H_0 &= - \sum_{ij\sigma} \sum_{ab} t^{ab}_{ij} \hat{c}_{ia\sigma}^\dagger c_{jb\sigma} - \mu \sum_{i a\sigma} n_{ia\sigma}\\
    H_\text{int} &= \frac{U_\text{ee}}{2} \sum_{i\sigma} \sum_{ab} n_{ia\sigma} n_{ib-\sigma} + \frac{U_\text{ee}}{2} \sum_{i\sigma} {\sum_{ab}}^\prime n_{ia\sigma} n_{ib\sigma}
\end{align}
where $t^{ab}_{ij}$ is effective hopping between site $j$ in orbital $b$ and site $i$ in orbital $a$, $U_\text{ee}$ characterizes the strength of  onsite Coulomb repulsion, and the prime in the summation of the latter term of $H_\text{int}$ indicates that the term $a = b$ is excluded.
The kinetic Hamiltonian can be diagonalized at each point in $k$-space as
\begin{equation}
    H_0 = \sum_{k\mu\sigma} \varepsilon_{\k\mu} c^\dagger_{\k\mu\sigma} c_{\k\mu\sigma}
\end{equation}
where $\mu$ denotes a \textit{band index}. 
For each $\k$, there exists $U^{a\mu}_\k$ such that $c_{\k a} = U^{a \mu}_\k c_{\k \mu}$. Expressing the interaction term in this basis, we obtain
\begin{equation}
    \begin{aligned}
        H_\text{int} &= \frac{U_\text{ee}}{N} \sum_{\k_1, \k_2, \q} \sum_{\sigma_1, \sigma_2} \sum_{\mu\nu\eta\tau} c^\dagger_{\k_1 - \q\eta\sigma_1}  F^{\eta\mu}_{\k_1 - \q, \k_1} c_{\k_1\mu\sigma_1} c^\dagger_{\k_2 + \q\tau\sigma_2} F^{\tau\nu}_{\k_2 + \q, \k_2} c_{\k_2\nu\sigma_2}
    \end{aligned}
\end{equation}
where $ F_{\k_1, \k_2}^{\mu\nu} \equiv \left ( U_{\k_1}^\dagger U_{\k_2} \right )^{\mu\nu}$. 

Since the interaction does not change spin, we can define the spinless quasiparticle interaction vertex as
\begin{align}
    W_{\mu_1, \mu_2, \mu_3, \mu_4}(\k_1, \k_2, \k_3, \k_4) &= \langle \k_1, \mu_1, \k_2, \mu_2| \hat{W} | \k_3, \mu_3, \k_4, \mu_4 \rangle\\
    &= U_\text{ee} F_{\k_3, \k_1}^{\mu_3\mu_1} F_{\k_4, \k_2}^{\mu_4\mu_2}.
\end{align}

As in \cite{Buhmann2013}, the scattering rates are taken to be given by Fermi's golden rule with an antisymmetrized scattering vertex to account for the anticommutativity of fermions as
\begin{multline}
\Gamma_{\mu_1\mu_2\mu_3\mu_4}^{\sigma_1 \sigma_2 \sigma_3 \sigma_4}(\k_1, \mu_1, \k_2, \mu_2, \k_3, \mu_3, \k_4, \mu_4) = \frac{2\pi}{\hbar} \delta(\varepsilon_{\mu_1}(\k_1) + \varepsilon_{\mu_2}(\k_2) - \varepsilon_{\mu_3}(\k_3) - \varepsilon_{\mu_4}(\k_4)) (2\pi)^2 \delta(\k_1 + \k_2 - \k_3 - \k_4)\\
\times \frac{1}{2} \left| W^{\sigma_1, \sigma_2, \sigma_3, \sigma_4}_{\mu_1, \mu_2, \mu_3, \mu_4}(\k_1, \k_2, \k_3, \k_4) - W^{\sigma_1, \sigma_2, \sigma_4, \sigma_3}_{\mu_1, \mu_2, \mu_4, \mu_3}(\k_1, \k_2, \k_4, \k_3)\right|^2 .
\end{multline}
Summing over spin polarizations and band indices, we can obtain an effective scattering rate
\begin{equation}
    \begin{aligned}
        \Gamma^\text{eff}(\k_1, \k_2, \k_3, \k_4) = \!\!\!\!&\sum_{\mu_1 \mu_2 \mu_3, \mu_4} 
        \left( \Gamma^{\sigma \sigma \sigma \sigma}_{\mu_1\mu_2\mu_3\mu_4}(\k_1, \k_2, \k_3, \k_4) + 2 \Gamma^{\sigma -\sigma -\sigma \sigma}_{\mu_1\mu_2\mu_3\mu_4}(\k_1, \k_2, \k_3, \k_4) \right)
    \end{aligned}
\end{equation}
which accords with an effective vertex
\begin{equation}
    \begin{aligned}
        W^2_\text{eff}(\k_1, \mu_1, \k_2, \mu_2, \k_3, \mu_3, \k_4, \mu_4) = U_\text{ee}^2 \left( |  F^{\mu_3,\mu_1}_{\k_3,\k_1} F^{\mu_4,\mu_2}_{\k_4,\k_2} -  F^{\mu_4,\mu_1}_{\k_4,\k_1} F^{\mu_3,\mu_2}_{\k_3,\k_2} |^2  + 2 |  F^{\mu_4,\mu_1}_{\k_4,\k_1} F^{\mu_3,\mu_2}_{\k_3,\k_2} |^2 \right)
    \end{aligned}
\end{equation}
such that
\begin{multline}
    \Gamma^\text{eff}(\k_1, \k_2, \k_3, \k_4) = \frac{\pi}{\hbar} (2\pi)^2\delta(\k_1 + \k_2 - \k_3 - \k_4)\\ \times \sum_{\mu_1 \mu_2 \mu_3, \mu_4} \delta(\varepsilon_{\mu_1}(\k_1) + \varepsilon_{\mu_2}(\k_2) - \varepsilon_{\mu_3}(\k_3) - \varepsilon_{\mu_4}(\k_4))\\
    \times W^2_\text{eff}(\k_1, \mu_1, \k_2, \mu_2, \k_3, \mu_3, \k_4, \mu_4).
\end{multline}
In practice, our calculations are performed at sufficiently low temperatures such that for each $\k$ there is an unique band index $\mu$ such that for all $\nu \neq \mu$,  $f^{(0)}(\varepsilon_\nu(\k)) (1 - f^{(0)}(\varepsilon_\nu(\k))) \ll 1$. 
Therefore, we will drop the sum over band indices, and the band index used for calculating energies will be implicit in $\k$. 
That is, the problem effectively reduces to the single band case, with the dispersion given as a piecewise function on implicit regions of the first Brillouin zone.

\section{Discretization of the Collision Operator \label{app:discretization}}
\begin{figure}[t!]
    \centering
    \includegraphics[width=0.5\textwidth]{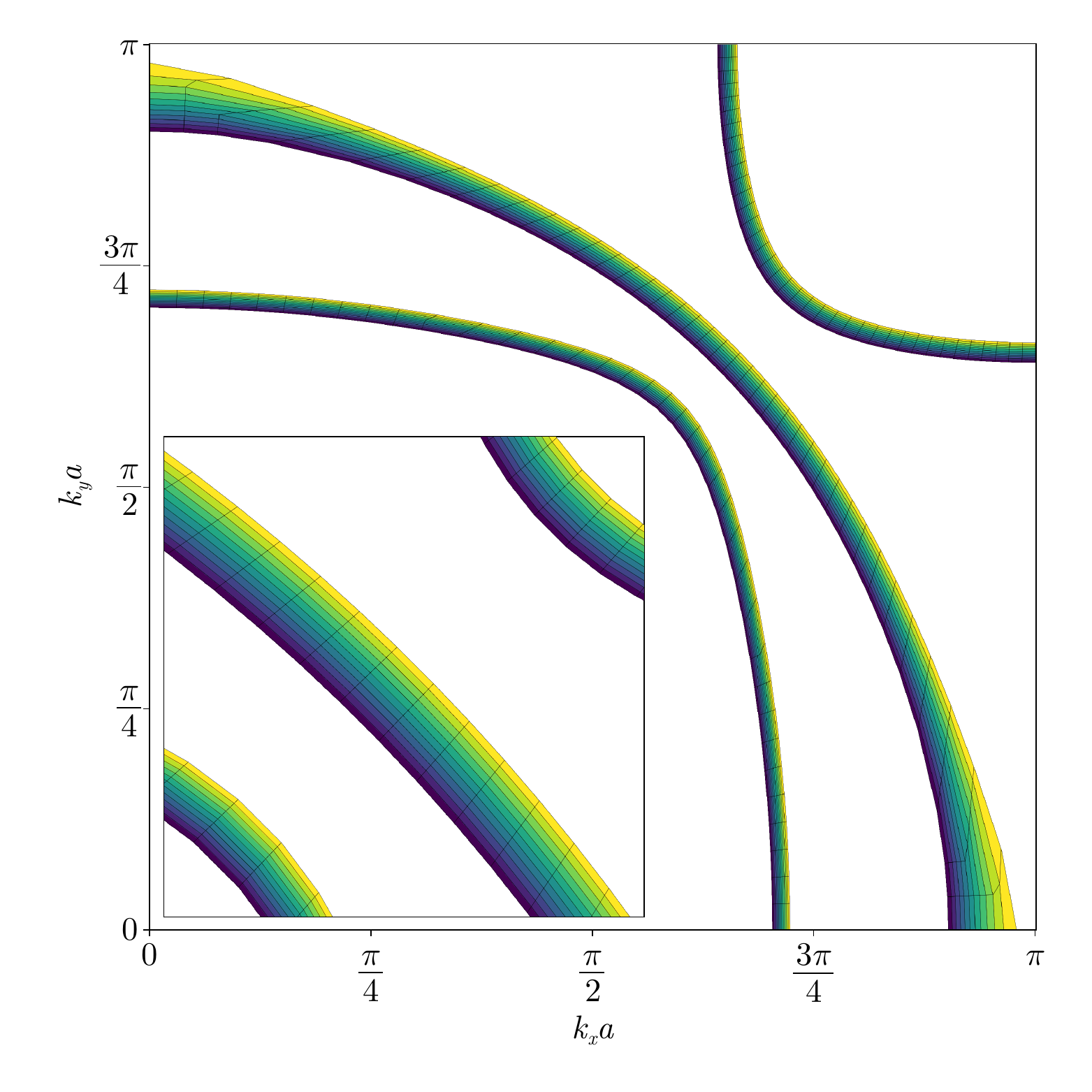}
    \caption{Fermi surface centered mesh of the first quadrant of the Brillouin zone at \SI{12}{\kelvin} at the sampled resolution. Each annular region has a width of $12 k_B T$. Inset: Zoom-in on the FSs along 110 to show the mesh.}
    \label{fig:mesh}
\end{figure}

Discretizing the Brillouin zone allows us to express the action of the collision integral upon an out-of-equilibrium distribution function as the following matrix equation.
\begin{equation}
    \begin{aligned}
        L[\chi(\k)]_i &= \sum_j \mathbf{L}_{ij} \chi_j\\
        \mathbf{L}_{ij} &= \frac{1}{dV_i} \int_i \frac{d^2\k_i}{(2\pi)^2} \int_j \frac{d^2\k_j}{(2\pi)^2} \mathcal{L}(\k_i, \k_j)
    \end{aligned}
\end{equation}
where $\int_i$ denotes an integral over the momenta contained in patch $i$ and  $dV_i$ is the area of patch $i$.
We adopt the discretization scheme of \cite{Buhmann2013} where patches are defined by binning momenta by energy and angle within an annular region of finite width following the Fermi surface of each band (see \autoref{fig:mesh}). While Umklapp scattering and impurity scattering relax momentum, particle number and total energy remain conserved quantities for which the corresponding functions $\chi_j = 1, \varepsilon_j$ are eigenvectors with vanishing eigenvalue. It follows that $\sum_{j} \mathbf{L}_{ij} = 0$, so that in practice only the off-diagonal elements are calculated, and the diagonal elements are obtained from this sum rule.

In terms of the effective scattering rate of \autoref{app:effective_rate}, the off-diagonal elements are given by

\begin{equation}
    \begin{aligned}
        \mathbf{L}_{ij} = \frac{1}{d V_i} \frac{\pi / \hbar}{1 - \fd_i} 
        & \frac{1}{(2\pi)^6} \sum_m \left( W^2_\text{eff}(\k_i, \k_j, \k_m, \k_i + \k_j - \k_m)  \fd_j (1 - \fd_m) \mathcal{K}_{ijm} \right . \\ & \left .
        - \left(W^2_\text{eff} (\k_i, \k_m, \k_j, \k_i + \k_m - \k_j) + W^2_\text{eff} (\k_i, \k_m, \k_i + \k_m - \k_j, \k_j) \right )\fd_m (1 - \fd_j) \mathcal{K}_{imj} \right)
    \end{aligned}
\end{equation}

with
\begin{equation}\label{eq:kernel}
        \mathcal{K}_{ijm} = \int_i d^2 \k_i \int_j d^2 \k_j \int_m d^2 \k_m (1 - \fd(\k_i + \k_j - \k_m)) \delta(\varepsilon(\k_i) + \varepsilon(\k_j) - \varepsilon(\k_m) - \varepsilon(\k_i + \k_j - \k_m))
\end{equation}
where all functions outside of patch integrals are approximated as constant within the patch and evaluated at the patch center.

Following \cite{Buhmann2013}, we make a coordinate transformation to energy-angle coordinates in each patch and linearize the argument of the delta function with respect to $\k_i, \k_j, \k_m$. The integral of \eqref{eq:kernel} can then be approximated as the volume of intersection of the hyperplane defined by the argument of the delta function with a hypersphere of equal volume to the integration volume, weighted by the value of the integrand on the hyperplane.

The resolution of the mesh is set by three parameters: the width of the annulus, the number $n_\varepsilon$ of energy bins, and the number $n_\theta$ of angular bins. All simulation data in the main text was generated with an annular width of $12 k_B T$, $n_\varepsilon = 11$, and $n_\theta = 148$ for each band.

\begin{figure}
    \centering
    \includegraphics[width=0.6\linewidth]{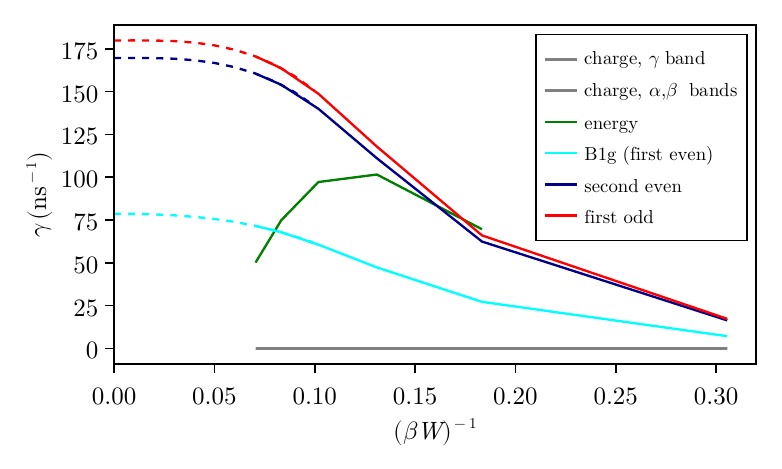}
    \caption{Convergence of the smallest eigenvalues of the collision operator at $T = 12$ K with the width $W$ of the Fermi annulus at constant sampling density in energy of $T^{-1}$ and $n_\theta = 148$. The charge modes on the two sets of bands have zero eigenvalues by construction. The eigenvalue labeled \Bog (first even) is the long-lived imbalance mode on $\alpha$ and $\beta$ discussed further below. The ``second even'' mode is the next parity-even mode when ordered by increasing eigenvalues. 
  The ``first odd'' mode is the first parity-odd mode (but turns out to be quite different from momentum since Umklapp is strong). Dashed lines serve as a guide to the eye and are obtained by cubic extrapolation with the convergence criterion $\lim\limits_{W \to \infty} d \gamma / d (\beta W)^{-1} = 0$.}
    \label{fig:convergence}
\end{figure}

Whereas conservation of particle number is enforced explicitly, the energy eigenvector acquires a small nonzero eigenvalue due to sampling on a finite-width energy window. 
\autoref{fig:convergence} demonstrates the convergence of the energy eigenvalue compared to those of the longest-lived modes for $\mathbf{L}_\text{ee}$. 
That the energy eigenvalue decreases while the other eigenvalues increase monotonically with width $W$ supports that energy will indeed be a zero mode in the limit where the entire Brillouin zone is sampled. 
For computing the conductivity and viscosity, the corresponding vector has zero overlap with the energy vector, and thus the nonzero energy eigenvalue makes no contribution to either transport coefficient.

\section{Electron-impurity scattering\label{app:imp}}
In the unitary limit, for an isotropic 2D band, the matrix element squared for electron-impurity scattering takes the form $| \langle \k_1 | V_\text{imp} | \k_2 \rangle |^2 = 4 \hbar^2 v_F^2 / k_F^2$. 
We define an analogous Fermi surface-averaged quantity for each band and treat interband scattering by taking the geometric mean of the weights,
\begin{equation}
    | \langle \k_1, \mu | V_\text{imp} | \k_2, \nu \rangle |^2 = 4 \sqrt{ \overline{\left(\frac{v_F^2}{k_F^2}\right)}_\mu \overline{\left(\frac{v_F^2}{k_F^2}\right)}_\nu }.
\end{equation}
We can also define the average impurity potential
\begin{equation}
    | \langle V_\text{imp} \rangle |^2_\text{av} = 4\sqrt[3]{\overline{\left(\frac{v_F^2}{k_F^2}\right)}_\alpha \overline{\left(\frac{v_F^2}{k_F^2}\right)}_\beta \overline{\left(\frac{v_F^2}{k_F^2}\right)}_\gamma }
\end{equation}
which serves as a single energy scale for characterizing the impurity strength.

The values of $v_F, k_F$ for each band from the tight-binding model of \autoref{app:tbm} are shown in \autoref{tab:fs-averages}.

\begin{table}[h!]
    \centering
    \begin{tabular}{c|c|c}
        Band & $\overline{v_F} \,(\si{\metre\per\second})$ & $\overline{k_F} \, (\si{\per\nano\metre})$ \\
        \hline
        $\alpha$ & 99395 & 2.97\\
        $\beta$ & 114577 & 6.14\\
        $\gamma$ & 59970 & 7.39
    \end{tabular}
    \caption{Fermi-surface-averaged Fermi velocities and Fermi momenta, calculated from the tight-binding model.}
    \label{tab:fs-averages}
\end{table}

\section{Tight-Binding Model\label{app:tbm}}
We use the nearest and next-nearest neighbor tight-binding approximation of the bands presented in \cite{BurganovEtAl2016} where
\begin{equation}
    H(\k) = \begin{pmatrix}
        \epsilon_{xz}(\k) & V(\k) & 0\\
        V(\k) & \epsilon_{yz}(\k) & 0\\
        0 & 0 &\epsilon_{xy}(\k) 
    \end{pmatrix}
\end{equation}
where
\begin{equation}
\begin{aligned}
    \epsilon_{xy}(\k) &= -\mu_1 - 2 t_1 \cos(k_x a) - 2 t_1 \cos(k_y a)\\
    &\quad - 4 t_4 \cos(k_x a)\cos(k_y a)\\
    \epsilon_{xz/yz}(\k) &= -\mu_1 - 2 t_2 \cos(k_{x/y} a) - 2 t_3 \cos(k_{x/y} a)\\
    V(\k) &= t_5 \sin(k_x a) \sin(k_y a).
\end{aligned}
\end{equation}
We take the values found in \cite{BurganovEtAl2016} for $t_4/t_1, t_3/t_2, t_5/t_2, \mu_1/t_1, \mu_2/t_2$ by fitting to the ARPES Fermi surface, but choose values of $t_1, t_2^\alpha, t_2^\beta$ which provide best agreement to the cyclotron mass for each band (see \autoref{tab:tbm}).

\begin{table}[h!]
    \centering
    \begin{tabular}{c|c|c|c|c|c|c|c}
         $t_1$ (eV) & $t_2^\alpha$ (eV)& $t_2^\beta$ (eV) & $t_4/t_1$ & $\mu_1/t$ & $t_3/t^{\alpha/\beta}_2$ & $t_5/t^{\alpha/\beta}_2$ & $\mu_2 / t^{\alpha/\beta}_2$\\
         \hline
         0.0774 & 0.099 & 0.128 & 0.392 & 1.48 & 0.08 & 0.13 & 1.08\\
    \end{tabular}
    \caption{Tight-binding model parameters. Ratios come from \cite{BurganovEtAl2016} with nearest-neighbor hoppings adjusted to provide agreement to the observed cyclotron mass for each band.}
    \label{tab:tbm}
\end{table}

\section{Deformation Potentials\label{app:deformation}}
We consider the band deformation potential of \cite{KhanEtAl1985} 
\begin{equation}
    D_{\alpha\beta}(\k, \mu) = \frac{\partial \varepsilon((\mathbf{1 - S})\k, \mu)}{\partial S_{\alpha\beta}} \Big \rvert_{\mathbf{S} = \mathbf{0}}
\end{equation}
where nearest and next-nearest neighbor hopping parameters of the tight-binding model are treated to linear order in strain as 
\begin{equation}
    \begin{aligned}
        t_{x/y} &= t_{x/y}^0(1 - \alpha S_{xx/yy})\\
        t^\prime &= (t^\prime)^0(1 - \alpha^\prime (S_{xx} + S_{yy})/2).
    \end{aligned}
\end{equation}
Here, $\alpha$ and $\alpha^\prime$ characterize the rate of deformation. Following \cite{NoadEtAl2023}, we take $\alpha = \alpha^\prime = 7.604$ which forces the $\gamma$ band to undergo a Lifschitz transition at a measured longitudinal strain of $S_{xx} = -0.44\%$ \cite{BarberEtAl2019}.

For sound attenuation in RUS, we are concerned with $\eta_{\Bog} = \eta_{xxxx} - \eta_{xxyy}$. Notably, $\eta_{xxxx} = \eta_{yyyy}$ and $\eta_{xxyy} = \eta_{yyxx}$ since the linearized collision operator is Hermitian under the inner product defined in the main text, allowing $\eta_{\Bog}$ to be written in the more symmetric form
\begin{equation}\label{eq:b1g}
    \begin{aligned}
        \eta_{\Bog} &= \frac{1}{4}(\eta_{xxxx} - 2\eta_{xxyy} + \eta_{yyyy})\\
        &= \left \langle \frac{D_{xx} - D_{yy}}{2} \Big |  L^{-1}\left[\frac{D_{xx} - D_{yy}}{2}\right] \right \rangle,
    \end{aligned}
\end{equation}
from which we can identify the deformation potential corresponding to \Bog strain as
\begin{equation}\label{eq:db1g}
    D_{\Bog} = \frac{1}{2}(D_{xx} - D_{yy}).
\end{equation}

\section{Heatmap of Electron-Electron Collision Operator, Hot Spots and Impact on Effective Lifetimes} 
\begin{figure}[h!]
    \centering
    \includegraphics[width=0.7\linewidth]{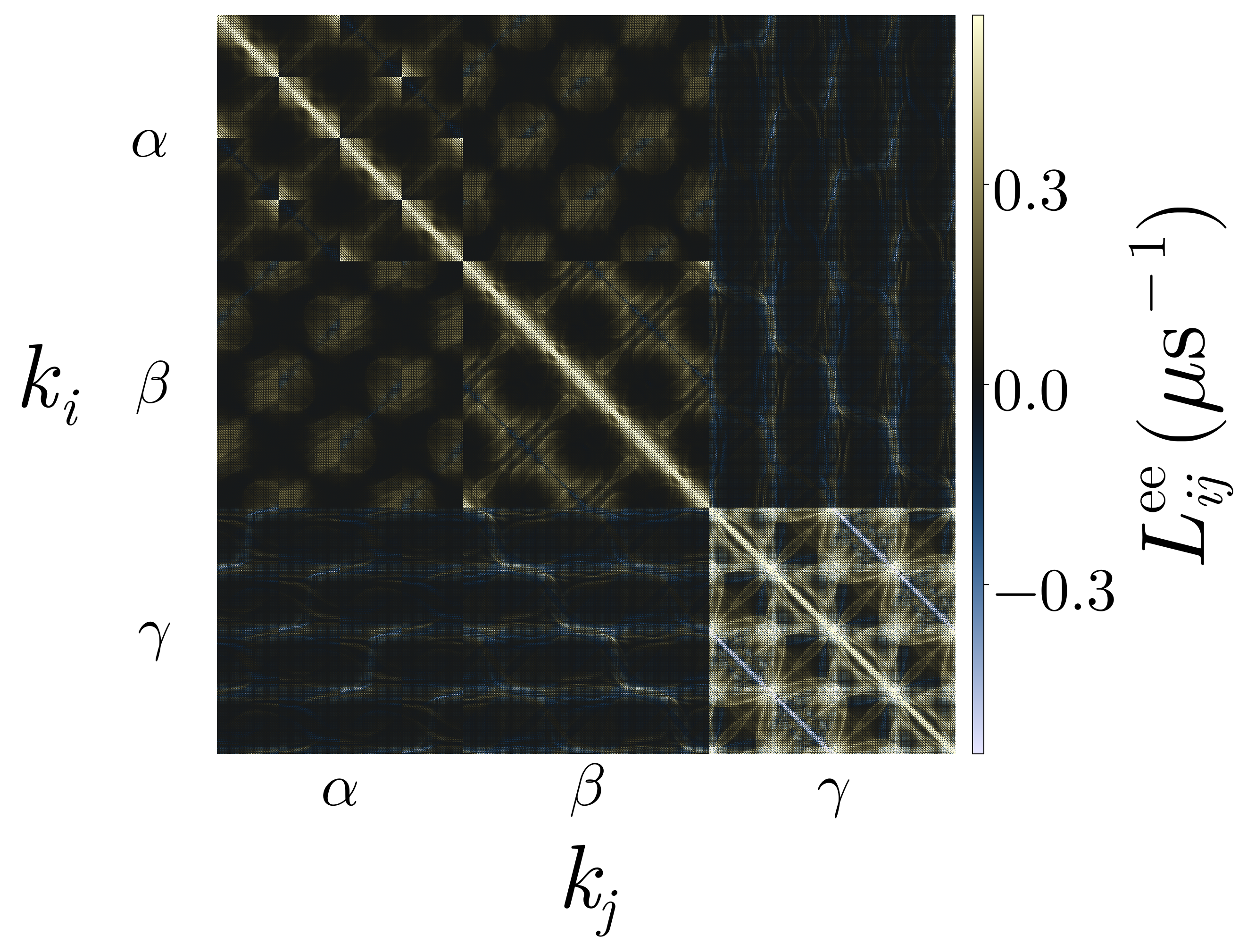}
    \caption{Heatmap of $L_\text{ee}$ at 12 K with resolution parameters $n_\varepsilon = 11$, $n_\theta = 148$. } 
    \label{fig:heatmap}
\end{figure}
\autoref{fig:heatmap} shows a heatmap of $L_\text{ee}$ at 12 K. The matrix can be decomposed into nine blocks, with the blocks along the main diagonal corresponding to \emph{intraband} scattering and the off-diagonal blocks corresponding to \emph{interband} scattering. The blocks are indexed such that $\alpha, \beta, \gamma \to 1,2,3$ for block row and column indices. Each block can further be decomposed into $n_\varepsilon \times n_\varepsilon$ blocks indexed by angle, which contain the matrix elements sampling over the energy direction in each corresponding angular slice. Angular slices start along 100 and increase counter-clockwise for $\beta$ and $\gamma$, whereas they start along 010 and increase clockwise for $\alpha$. See Fig.~\ref{fig:mesh} for a plot of our mesh.

\begin{figure}[h!]
    \centering
    \includegraphics[width=0.6\linewidth]{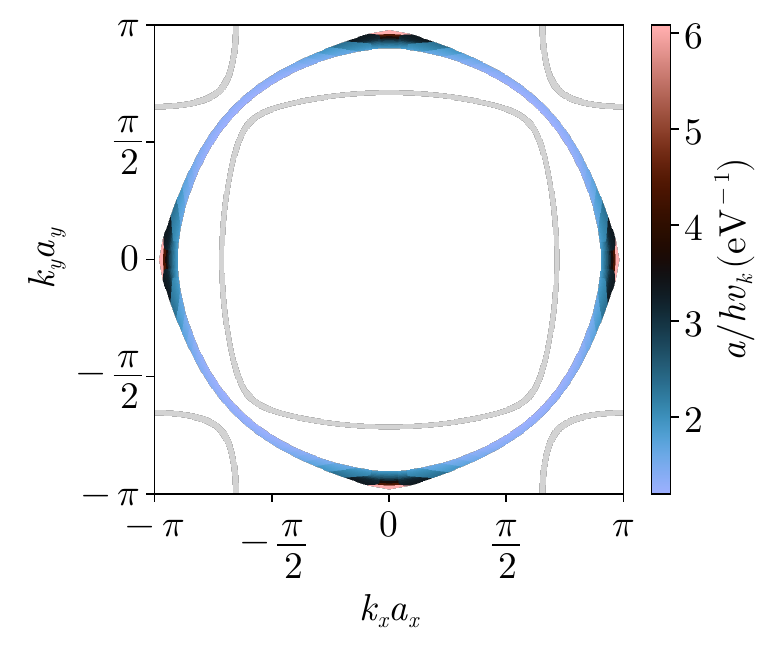}
    \caption{Inverse group velocity highlighted for the $\gamma$ band. Hot spots exist along the (100) and (010) directions where the $\gamma$ band approaches the zone boundary.}
    \label{fig:hot-spots}
\end{figure}

In each diagonal block, there are two prominent features: (1) a forward scattering peak along the main diagonal, and (2) a backscattering depression located at $\pm \pi$ from the main diagonal. Both these features are expected for two-dimensional electron-electron scattering and were reported for circular Fermi surfaces in the literature~\cite{LedwithEtAl2019,HongEtAl2020}.
However a rich structure of additional features reflecting the Fermiology of \SRO can also be seen in the collision operator.
The $(\gamma, \gamma)$ block features bright lines regularly spaced at intervals of $\pi/2$. These correspond to ``hot spots'' where the density of states is enhanced in the vicinity of the van Hove singularity (see \autoref{fig:hot-spots}). 

As mentioned in the main text, these hot spots are responsible for the 30\% difference between the $T^2$ prefactors of the inverse lifetimes $\tau_\sigma^{-1}$ and $\tau_{\eta}^{-1}$ giving the conductivity and \Bog viscosity respectively.
In short, the 30\% larger prefactor for $\tau_\eta^{-1}$ is due to the fact that scattering of an electron between two hot spots separated by an angle of $\pm \pi/2$ provides a fast way to relax $B_{1g}$, whereas the role of hot spots for current relaxation is not so direct, as we now explain.
On the one hand, hot spots host enhanced Umklapp scattering since they allow small momentum scattering across the zone boundary.
On the other hand, Umklapp scattering is much weaker for electrons at ``cold spots'' along $(110)$ and $(1\bar{1}0)$.
These electrons thus need to first scatter to a hot spot before undergoing Umklapp scattering there.
We find that this cold to hot spot scattering is comparatively slow and thus acts as a bottleneck, eventually leading to a smaller $\tau_\sigma^{-1}$. 
Interestingly, scattering between cold and hot spots was already proposed as a bottleneck for transport in \SRO brought in close proximity to a Lifshitz transition using uniaxial strain~\cite{Mousatov}.

We should also comment on the low temperature behavior of the effective lifetimes.
Fig.~2c of the main text shows that at low-$T$ the ratio of scattering rates is inverted: the effective scattering rate for the viscosity is smaller than that for the conductivity. This is due to a multi-band effect: at low-$T$, electron-impurity scattering dominates. In the unitary limit, this leads to a smaller scattering rate on the $\gamma$ band due to its higher effective mass. Since the weight on the $\gamma$ band is comparatively higher for the deformation potential $|D_{\Bog}\rangle$ than for the current $|v_x\rangle$, this leads to a comparatively smaller $\tau_\eta^{-1}$ in the impurity-dominated regime at low temperature.

\section{Nonlocal Conductivity for $\gamma$ band only model}
\begin{figure}[h!]
    \centering
    \includegraphics[width=0.5\linewidth]{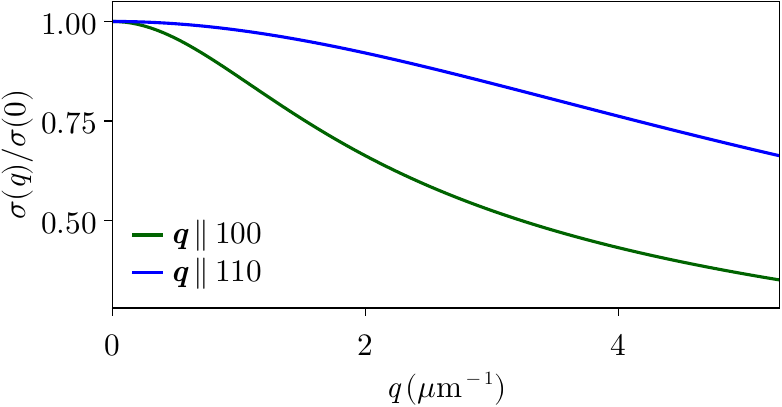}
    \caption{Non-local conductivity calculated at $T = 14$ K for the $\gamma$ band only model.
    }
    \label{fig:nonlocal_gamma}
\end{figure}

In this appendix we discuss results for a single-band model where only the $\gamma$ band is included, in order to contrast it with our results in the main text obtained for a three-band model.
We perform a new fit to the resistivity for this single-band model and find $U_{\text{ee}, \gamma} = \SI{0.040}{\electronvolt}$ and $V_{\text{imp}, \gamma} = \SI{2.8e-5}{\electronvolt}$. The nonlocal conductivity for this model is shown in \autoref{fig:nonlocal_gamma} and exhibits an anisotropy with 110 as the easy axis (i.e. $\sigma(q_{110})>\sigma(q_{100})$).
Since this anisotropy is opposite to the one we found in the full three-band calculation reported in the main text (for which 100 was the easy axis), we conclude that $\alpha$ and $\beta$ dominate non-local transport and are responsible for the anisotropy in the three-band model.

\section{Imbalance Mode}
\begin{figure}
    \centering
    \includegraphics[width=0.6\linewidth]{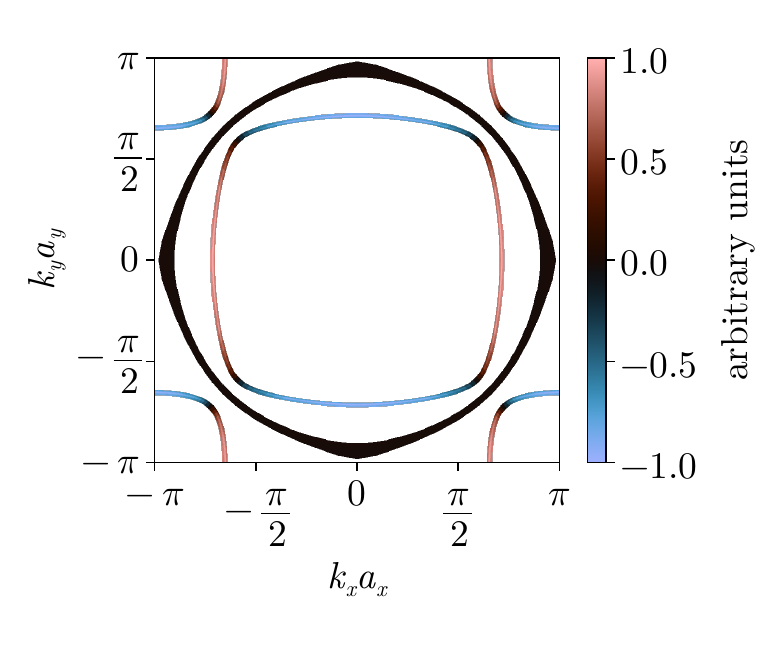}
    \caption{Eigenvector of the collision operator corresponding to a long-lived imbalance mode. The geometry of the $\alpha$ and $\beta$ Fermi surfaces is responsible for restricting the available phase space for scattering events that can relax this kind of perturbation.
    }
    \label{fig:imbalance-mode}
\end{figure}

The longest-lived eigenmode of $L_\text{ee}$ (apart from density and energy) is found to exhibit a \Bog modulation over the $\alpha$ and $\beta$ bands with negligible weight on the $\gamma$ band (see \autoref{fig:imbalance-mode}). Microscopically, it corresponds to an imbalance between the occupation of $d_{zx}$ and $d_{zy}$ orbitals. This type of perturbation can only relax through scattering from positive (red) to negative (blue) regions. As discussed in \citet{CookLucas}, the available phase space for scattering between adjacent edges on the perfectly square Fermi surface is restricted, leading to a comparatively small scattering rate. For the square-like Fermi surfaces of the $\alpha$ and $\beta$ bands, this effect is preserved such that the imbalance mode shown in \autoref{fig:imbalance-mode} has a long relaxation time.

\clearpage

\bibliography{main.bib}